\documentclass[prl,twocolumn,floatfix,superscriptaddress,amsmath,citeautoscript,aps,longbibliography]{revtex4-2}
\pdfoutput=1
\usepackage[T1]{fontenc}\usepackage[latin1]{inputenc}
\usepackage{dcolumn,bm,graphicx,color,booktabs,microtype,afterpage} \graphicspath{{Figures/}}
\usepackage[charter,greekuppercase=italicized]{mathdesign}
\renewcommand{\figurename}{Fig.}
\renewcommand{\tablename}{Table}
\makeatletter\renewcommand{\fnum@figure}[1]{\figurename~\thefigure.}\makeatother
\makeatletter\renewcommand{\fnum@table}[1]{\tablename~\thetable.}\makeatother
\newcount\hh \newcount\mm
\hh=\time \divide\hh by 60
\mm=\hh \multiply\mm by 60 \mm=-\mm
\advance\mm by \time
\def\now{\number\hh:\ifnum\mm<10{}0\fi\number\mm}
\usepackage[colorlinks,plainpages=false,linkcolor=blue,urlcolor=blue,citecolor=blue,pdfpagemode=UseNone,pdfstartview=FitBH]{hyperref}

\newcommand{\dfo}{DyFeO$_3$}

\begin{document}

\title{Competition Between Multiferroic and Magnetic Soliton Lattice States in \dfo}

\author{S.~E.~Nikitin}
\affiliation{PSI Center for Neutron and Muon Sciences, Paul Scherrer Institut, CH-5232 Villigen-PSI, Switzerland}
\thanks{stanislav.nikitin@psi.ch}
\author{N.~D.~Andriushin}
\affiliation{Institut f{\"u}r Festk{\"o}rper- und Materialphysik, Technische Universit{\"a}t Dresden, D-01069 Dresden, Germany}
\author{{\O}.~S.~Fjellv{\aa}g}
\affiliation{PSI Center for Neutron and Muon Sciences, Paul Scherrer Institut, CH-5232 Villigen-PSI, Switzerland}
\affiliation{Department for Hydrogen Technology, Institute for Energy Technology, Kjeller, NO-2027, Norway}
\author{E.~Pomjakushina}
\affiliation{PSI Center for Neutron and Muon Sciences, Paul Scherrer Institut, CH-5232 Villigen-PSI, Switzerland}
\author{A.~A.~Turrini}
\affiliation{PSI Center for Neutron and Muon Sciences, Paul Scherrer Institut, CH-5232 Villigen-PSI, Switzerland}
\author{S.~Artyukhin}
\affiliation{Quantum Materials Theory, Italian Institute of Technology, Via Morego 30, 16163 Genova, Italy}
\author{C.~W.~Schneider}
\affiliation{PSI Center for Neutron and Muon Sciences, Paul Scherrer Institut, CH-5232 Villigen-PSI, Switzerland}
\author{M.~Mostovoy}
\affiliation{Zernike Institute for Advanced Materials, University of Groningen, Nijenborgh 4, 9747 AG Groningen, The Netherlands}

\begin{abstract}

Simultaneous breaking of time reversal and inversion symmetries in multiferroics couples ferroelectricity to magnetism and is a source of unusual physical phenomena that can be used in next-generation electronic devices.
A notable example is DyFeO$_3$, which under applied magnetic fields exhibits a giant linear magnetoelectric response and a large spontaneous electric polarization induced by coexisting orders of Fe and Dy spins. 
Here, we use high-resolution neutron diffraction to show that at zero field \dfo\ hosts an incommensurate magnetic soliton lattice formed by spatially ordered Dy domain walls with an average domain size of 231(8)~\AA.
The long-ranged interaction between the domain walls is mediated by magnons propagating through the Fe subsystem and is analogous to the Yukawa force in particle physics. 
An applied  magnetic field destroys the long-ranged incommensurate order, unlocks the linear magnetoelectric response and stabilizes the ferroelectric state. 
The magnetic domain walls are electrically charged and the soliton array dimerizes when both electric and magnetic fields are applied. 
Numerical simulations with experimental parameters suggest, that the generic competition between the ferroelectric and incommensurate states can be effectively controlled by an applied electric field.

\end{abstract}
\maketitle

\textbf{Introduction}. 
Multiferroics are fascinating materials in which the interplay between ferroelectricity and magnetism presents exciting opportunities for advanced applications in memory devices, sensors, and spintronics~\cite{tokura2014multiferroics, fiebig2016evolution, spaldin2019advances, mostovoy2024multiferroics}. Many multiferroics are frustrated magnets showing unconventional magnetic orders, e.g. spin-spirals~\cite{kimura2003magnetic, kenzelmann2005magnetic, arkenbout2006ferroelectricity} and $\uparrow\uparrow\downarrow\downarrow$-state~\cite{choi2008ferroelectricity, yanez2011multiferroic}, which spontaneously break inversion symmetry and induce ferroelectricity. Frustration can enable the coexistence of two magnetic orders, $\Gamma$ and $\Gamma'$, one even and another odd under inversion. Together, they can give rise to rather high electric polarization, e.g. $P = 0.35$ $\frac{\rm \mu C}{{\rm cm}^2}$ measured in GdMn$_2$O$_5$ with two types of antiferromagnetically ordered chains~\cite{lee2013giant, ponet2022topologically}, and $P = 0.1$ $\frac{\rm \mu C}{{\rm cm}^2}$ in the orthoferrite GdFeO$_3$ with interpenetrating sublattices of Fe and rare-earth ions~\cite{tokunaga2009composite}. Another orthoferrite, TbFeO$_3$, is paraelectric and shows an unconventional incommensurate (IC) soliton array state that involves both Fe and Tb spins~\cite{artyukhin2012solitonic}. The competition between the ferroelectric and IC states in orthoferrites is not accidental: When magnetoelectric coupling, $P\Gamma \Gamma'$ is allowed by symmetry, so is the Lifshitz invariant, $\Gamma \overleftrightarrow{\partial} \Gamma' = \Gamma \partial \Gamma' - \Gamma' \partial \Gamma$, $\partial$ being the derivative along the polarization direction, which favors periodically modulated magnetic orders in {\em centrosymmetric} magnets.

In this work, we use a combination of single-crystal neutron diffraction and theoretical analysis to study the close competition of the IC and ferroelectric (FE) states in \dfo. We show that in zero field Dy domain walls (DWs) form a periodic array stabilized by long-ranged interactions mediated by magnons, similar to that observed in TbFeO$_3$~\cite{artyukhin2012solitonic}. The application of a magnetic field along the $c$-axis suppresses the IC state in favor of a state with a short-range commensurate order at $B \approx\ 2.6$~T. Due to magnetoelectric coupling, the IC state can also be suppressed by applying an electric field. In zero magnetic field, the critical electric field is rather high, but it decreases rapidly under applied magnetic fields, making the electric control of magnetism in DyFeO$_3$ feasible.

\begin{figure*}[tb]
\center{\includegraphics[width=1\linewidth]{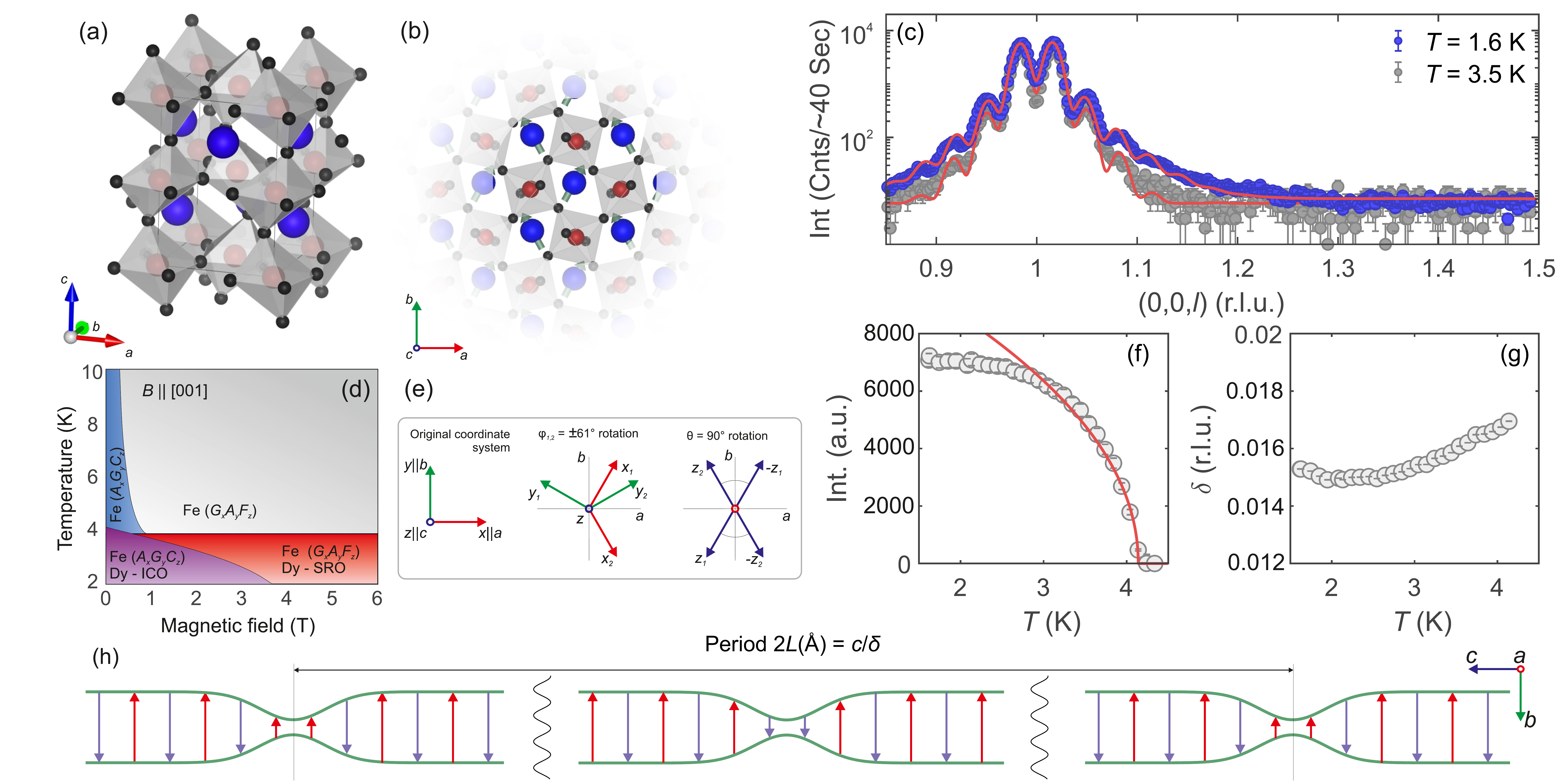}}
  \caption{~\textbf{Formation of magnetic soliton lattice in \dfo} 
  (a,b)~Crystal structure of \dfo. Blue, red and black balls represent Dy, Fe and O ions, respectively. Green arrows in panel (b) demonstrate the ordering pattern of Dy moments in the $ab$-layer at zero field.
  (c)~Neutron diffraction data measured along the $(0,0,l)$ direction at $T = 1.7$ and 3.5~K. The intensity is shown on a logarithmic scale. Red lines show the fitting results with the Monte-Carlo model as described in the main text. The absorption correction was included in the calculated curve.
  (d)~Schematic field-temperature phase diagram of \dfo{}~\cite{wang2016simultaneous} for $B \| c$. The blue and grey area correspond to $\Gamma_1$ and $\Gamma_4$ magnetic orders of the Fe subsystem, and the disordered state within the Dy subsystem. The violet area corresponds to the IC Dy order. The application of a magnetic field causes a spin reorientation of Fe moments, $\Gamma_1\rightarrow\Gamma_4$ and stabilizes a short-range commensurate Dy order (red area).
  (e)~Visualization of the Ising axis directions of Dy$^{3+}$ moments. The axes lie within the $ab$-plane at $\pm29^{\circ}$ to the $b$-axis.
  (f,g)~Temperature dependence of intensity (f) and IC parameter $\delta$ of the Dy order. 
  (h)~Sketch of the incommensurate magnetic structure of the Dy subsystem. Dy moments form an AFM structure with a domain size $L = \frac{c}{2\delta}$.
  }
  \label{Fig1}
  \vspace{-12pt}
\end{figure*}

\textbf{Single-ion physics of Dy$^{3+}$ and the magnetic structure of \dfo}.
Rare-earth orthoferrites $R$FeO$_3$ host two interpenetrating magnetic sublattices [Fig.~\ref{Fig1}(a,b)] and the interaction between them results in a spontaneous spin-reorientation transition (SRT)~\cite{white1969review, belov1976spin}, ultrafast laser-induced magnetization switching~\cite{kimel2004laser, mikhaylovskiy2014terahertz}, multimagnon spin excitations~\cite{Nikitin2018} and multiferroicity that appears together with the giant linear magnetoelectric response ~\cite{tokunaga2008magnetic, tokunaga2009composite}. Magnetism in these compounds is determined by strongly hierarchical interactions. The behavior of the Fe-subsystem is governed by the dominant Fe-Fe exchange coupling that favors a simple G-type AFM order ($\mathbf{k} = 0$) and a much weaker Dzyaloshinskii-Moriya interaction that cants spins and produces a weak ferromagnetic (FM) moment~\cite{Podlesnyak2021}. A weak magnetic anisotropy of Fe ions orients the N\'eel vector $\mathbf{n}$ along the $c$-axis and the resulting magnetic state is described by the $\Gamma_4$ irreducible representation (irrep).

With decreasing temperature, $\mathbf{n}$ reorients toward the $b$-axis ($\Gamma_1$ irrep) at $T_{\rm SR} = 40\,-\,50$~K, while the Dy spins order below $\sim 4$~K~\cite{ritter2022magnetic}. However, there are controversies regarding the Dy order: single-crystal neutron diffraction experiments yielded a commensurate order with $\mathbf{k}_{\rm m} = (0,0,1)$ ($\Gamma_5$ irrep)~\cite{wang2016simultaneous}. Such symmetry allows for a $\Gamma_4 \overset{\leftrightarrow}{\frac{\partial}{\partial{}z}}\Gamma_5$ Lifshitz invariant, that favors an IC modulation along the $c$-axis and could suppress the electric polarization. Indeed, recent high-resolution powder diffraction experiments showed a small splitting of the (0,0,1) peak indicative of IC modulations along the $c$-axis, $\mathbf{k}_{\rm m} = (0,0,1\pm\delta)$~\cite{ritter2022magnetic, biswas2022role}, but the refinement of the powder diffraction data did not allow to uniquely identify the ordering pattern of the Dy spins suggesting either a collinear spin-density wave (SDW) or an elliptical spiral as the ground state.


\begin{figure}[tb]
\center{\includegraphics[width=1\linewidth]{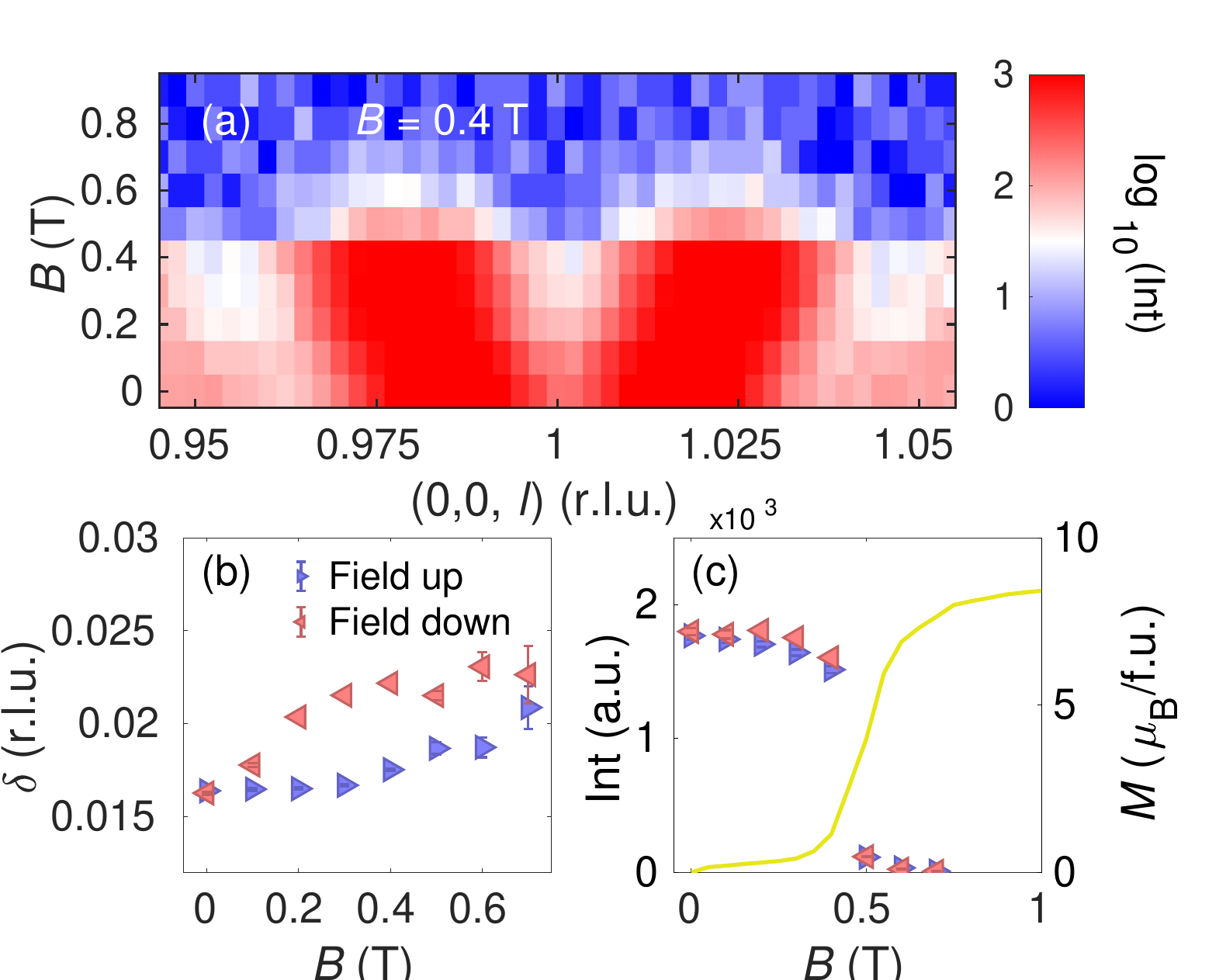}}
  \caption{
    \textbf{Field-induced ferromagnetic state.}
    ~(a)~Field dependence of the neutron intensity measured along $(0,0,l)$ at $T = 1.7$~K with the magnetic field applied along the $b$-axis (easy axis of magnetization). The intensity is shown on a logarithmic scale.
    (b,c)~Field dependence of the $\delta$ parameter (b) and peak intensity (c) obtained with the magnetic field ramped up (blue) and down (red). 
    Yellow line in panel (c) demonstrates bulk magnetization from Ref.~\cite{zhao2014ground} measured at $T = 2$~K.   }
  \label{Fig_010}
  \vspace{-12pt}
\end{figure}

\begin{figure*}[tb]
\center{\includegraphics[width=1\linewidth]{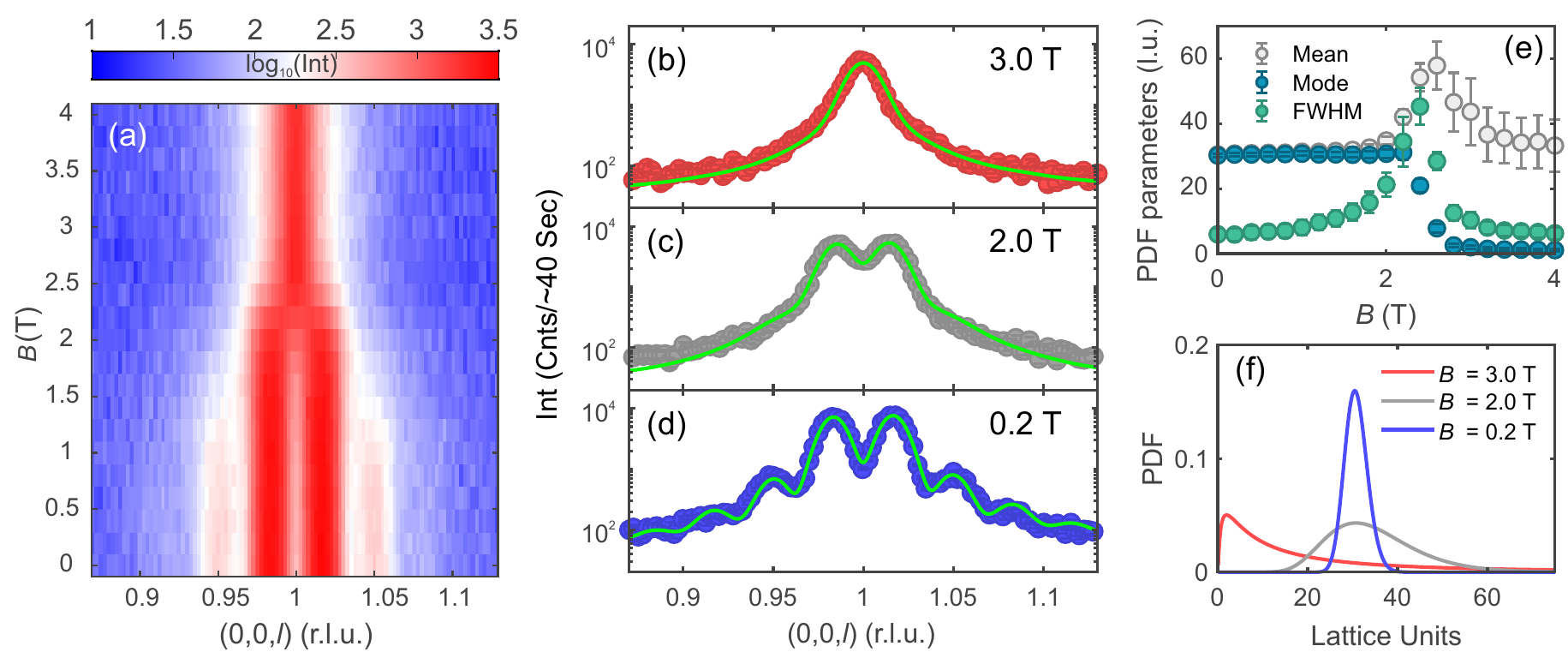}}
  \caption{
    \textbf{Field-induced suppression of the soliton lattice and domain wall distribution.}
    (a)~Field dependence of the neutron intensity measured along $(0,0,l)$ at $T = 1.7$~K. The intensity is shown on a logarithmic scale with the magnetic field applied along the $c$-axis.
    (b-d)~Representative scans along the $(0,0,l)$ direction taken at 3~T (b), 2~T (c) and 0.2~T (d) (dots represent the measured signal). The green lines are the absorption corrected fits to the data using the Monte Carlo model with the intensity shown on a logarithmic scale (Sec.~xx of SM~\cite{SM}).
    (e)~Mode, mean and full width half maximum (FWHM) of the pair distribution function (PDF)  fitted for different fields.
    (f)~Representative PDF extracted from our model of the data shown in panels (b-d).
  }
  \label{Fig2}
  \vspace{-12pt}
\end{figure*}

The AFM ordering of the $R$ ions in orthorhombic $RM$O$_3$ is primarily determined by a crystal electric field (CEF)-induced anisotropy and ranges  between a collinear $c$-axis structure (CeFeO$_3$~\cite{ritter2021determination}, ErFeO$_3$~\cite{deng2015magnetic}) and non-collinear orders when the rare-earth moments lie within the $ab$ plane (TbFeO$_3$~\cite{artyukhin2012solitonic}, HoFeO$_3$~\cite{chatterji2017temperature}, DyFeO$_3$~\cite{ritter2022magnetic, biswas2022role}). 
Point-charge model (PCM) calculations supported by experimental data for the isostructural DyScO$_3$ demonstrated that Dy moments have a very strong Ising-like anisotropy with the spin-quantization axes lying within the $ab$-plane, $\approx \pm 28^{\circ}$ from the $b$-axis~\cite{Wu2017DyScO3, andriushin2022slow}. We analyzed the ground state of Dy$^{3+}$ ions in \dfo\ using PCM and found a very similar anisotropy (see details in Sec.~S1 of SM~\cite{SM}) with a canting angle of $\approx \pm 29^{\circ}$, see Fig.~\ref{Fig1}(e). The ground state doublet, $|\psi_{\pm}> \approx\ |\pm 15/2>$, has only a negligible (<1~\%) admixture of other states, and thus exhibits a very strong Ising-like anisotropy in agreement with previous analysis~\cite{zvezdin1979theory}. 

This result is important in the context of magnetic structure determination. The strongly anisotropic Ising-like Dy moments make the spin-spiral implausible because it involves a continuous rotation of the moments within the $ab$-plane. The second solution yields two SDW with different magnetization directions. Both lie within the $ab$-plane, rotated by $\pm 29^{\circ}$ with respect to the $b$-axis~\cite{ritter2022magnetic}. The magnetization direction perfectly coincides with the direction of the Ising axis obtained from the PCM calculations, providing strong support for the SDW-like scenario. However, the general-type SDW involving a continuous modulation of the moment, is also unlikely for the Ising doublet with $\psi_{\pm} \approx |\pm15/2>$. Instead, incommensurability in Ising-class models [such as an anisotropic next-nearest neighbor Ising (ANNNI) model] can arise from the spatial ordering of the DWs~\cite{selke1988annni} similar to the one observed in TbFeO$_3$~\cite{artyukhin2012solitonic}.

\textbf{Magnetic structure studied by neutron scattering}.
We studied the field-induced evolution of magnetic order in \dfo\ by means of high-resolution single-crystal neutron diffraction using the TASP instrument at PSI. We focused primarily on the close vicinity of the $(0,0,1)$ peak, which is the strongest magnetic reflection of the Dy subsystem~\cite{ritter2022magnetic}. Figure~\ref{Fig1}(c) shows the representative diffraction profiles measured at zero field and base temperature. It exhibits two strong incommensurate peaks at $\mathbf{k}_{\rm m} = (0,0,1\pm\delta)$, four additional satellites at $3\mathbf{k}_{\rm m}$ and $5\mathbf{k}_{\rm m}$ and broad diffuse scattering extending up to $\approx{}1.3$~r.l.u.. The even order satellites at $2n\mathbf{k}_{\rm m}$ have not been observed at zero field~\footnote{Due to the dark space of the horizontal magnet, we could not access the $\mathbf{Q}$ position of the even-order satellites at finite fields.}.
The IC parameter $\delta = 0.01646(8)$ slightly differs from the value obtained from powder measurements by Ritter \textit{et al.} $\delta = 0.026-0.028$~\cite{ritter2022magnetic} but agrees nicely with $\delta =  0.0173(5)$ obtained by Biswas \textit{et al.}~\cite{biswas2022role}. Such a difference indicates a sample-dependence of the propagation wavevector. $\delta$ allows us to directly estimate the real space periodicity of the magnetic structure, $c/(\delta) = 462(16)$~\AA\ ($c$ is the lattice parameter). With increasing temperature, the intensity of the AFM peak is continuously suppressed, eventually disappearing at $T_{\rm N} = 4.14(2)$~K, see Fig.~\ref{Fig1}(f). The IC parameter $\delta$ exhibits a minor increase between 1.6~K and $T_{\rm N}$ [Fig.~\ref{Fig1}(g)].

Next, we looked into the field-induced evolution of the magnetic order. When a magnetic field is applied along the easy axis, $B\|b$, it quickly polarizes Dy moments and induces a FM state, as evident from a sharp jump and quick saturation of the magnetization at $B \geq\ 0.5$~T [Fig.~\ref{Fig_010}(c)]. Neutron data yield a strong suppression of the AFM reflections at the same magnetic field, as demonstrated in Fig.~\ref{Fig_010}(a,c). The parameter $\delta$ exhibits a moderate enhancement with field for the reason discussed below and shows hysteretic behavior, likely related to the DW pinning.

\begin{figure*}[tb]
\center{\includegraphics[width=1\linewidth]{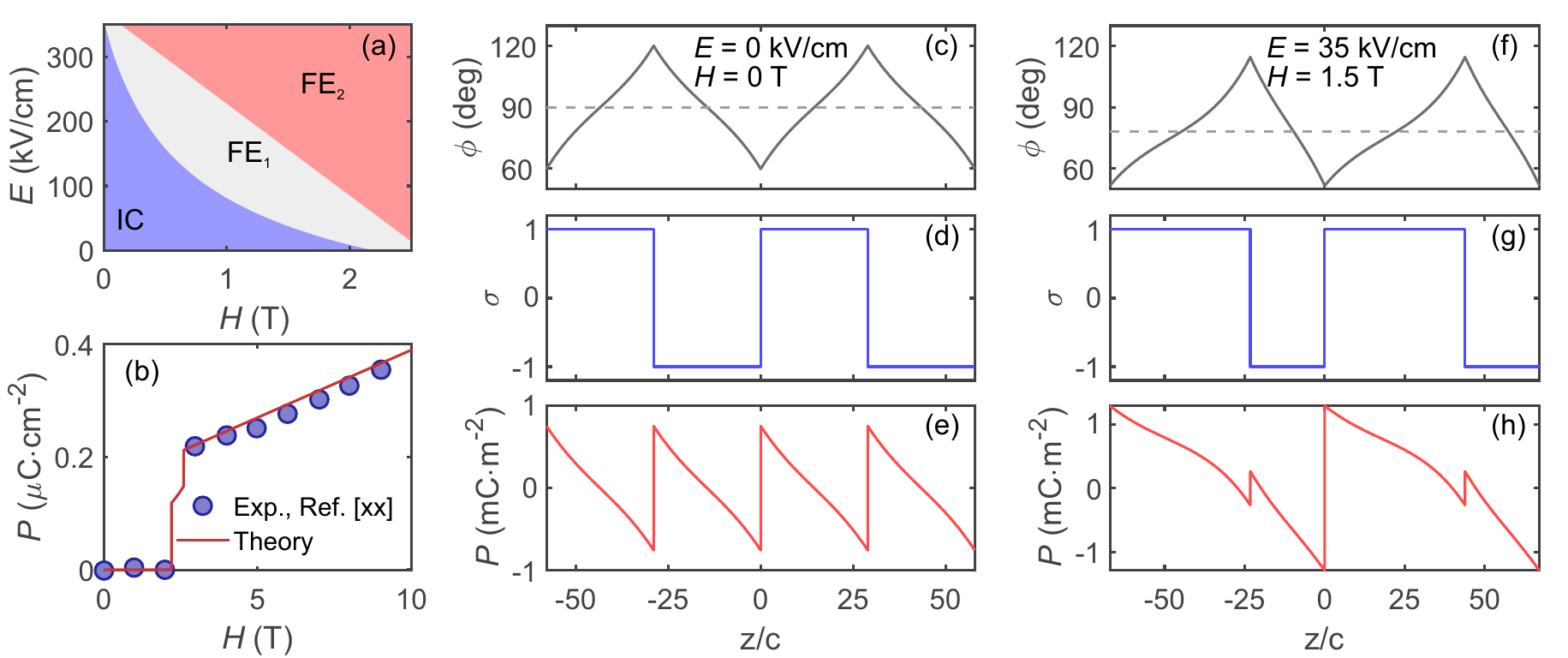}}
  \caption{
    \textbf{Origin of the incommensurate order and the phase diagram of \dfo}
    (a)~Three phases obtained by numerical minimization of the free energy density in applied electric and magnetic fields: the IC state (blue), the uniform ferroelectric state, FE$_1$ with $\sigma = 1$ and $0 < \phi < 90^{\circ}$ (grey), and the uniform ferroelectric state FE$_2$ (red) with Dy spins in the $\Gamma_5$-state and $\phi = 0$ ($\Gamma_4$ symmetry).
    (b)~Magnetic field dependence of the electric polarization from Ref.~\cite{tokunaga2008magnetic} and calculated by our theory.
    (c-h) Coordinate dependence of the angle $\phi$ describing the direction of the N\'eel vector in the IC state (c,f), the sign change of the Dy order parameter $\sigma$ (d,g) and the electric polarization (e,h). 
    Data are calculated for $H_0 = E_0 = 0$ (c-e) and $H_0 = 1.5$~T, $E_0 = 35 $kV$\cdot$cm$^{-1}$ (f-h). The dashed line in the panels (c,f) show the average value of $\phi$. Distances are measured in units of the lattice constant $c$.
  }
  \label{Fig3}
  \vspace{-12pt}
\end{figure*}

The  magnetic field applied along the $c$-axis 
considerably modifies the observed diffraction profile [see  Fig.~\ref{Fig2}(a)]. We can broadly categorize the field evolution into three regimes: (i) at $B < 2.2$~T the positions of the primary peaks remain constant, but the higher-order satellites continuously decrease in intensity forming a broad diffusive tail; (ii) at $2.4 <B < 2.6$~T two primary IC satellites merge into a commensurate peak at (0,0,1); (iii) at $B > 2.6$~T a single Lorentzian peak centered at $(0,0,1)$ dominates the pattern. Three representative patterns taken at 3, 2 and 0.2~T are shown in Figs.~\ref{Fig3}(b-d). 
 
Thus, our single crystal data clearly support the formation of the IC state observed in powder diffraction experiment at zero field~\cite{ritter2022magnetic, biswas2022role}. A magnetic field applied along an in-plane direction directly couples to the Ising-like Dy moments and quickly polarizes the Dy subsystem as shown in Fig.~\ref{Fig_010} for $B\|b$ and in Sec.~S2 of SM~\cite{SM} for $B\|[110]$. In contrast, a magnetic field applied along the $c$-axis (hard axis of Dy moments) primarily interacts with the Fe subsystem, causing the $\Gamma_1\xrightarrow\ \Gamma_4$ transition. However, $B\|c$ indirectly controls Dy order via the Dy-Fe coupling. As we show next, the AFM order of Dy spins does not change locally, but rather suppresses the IC long-range ordering of the DWs in favor of the commensurate AFM state with randomly positioned DWs. Ferroelectricity and a strong linear magnetoelectric response are observed in the commensurate phase~\cite{tokunaga2008magnetic}.

\textbf{Monte-Carlo simulations of neutron diffraction data}. 
In view of the Ising nature of Dy moments, the most plausible picture of the IC state is the spatial ordering of Dy DWs similar to the one observed in TbFeO$_3$~\cite{artyukhin2012solitonic}. These walls separate the domains with opposite signs of the AFM Dy spin order and form a magnetic soliton lattice [Fig.~\ref{Fig1}(h)]. Following this idea, we describe our neutron scattering data using an empirical Monte-Carlo approach. 
This allows to extract the average distance between the domain walls and quantify the degree of disorder. We consider an AFM Ising spin chain as schematically demonstrated in Fig.~\ref{Fig1}(h) and assume that the distances between domain walls (i.e. domain sizes) follow a log-normal distribution. 
We further assume that the DWs in the Dy order have a small but finite width due to a suppression of ordered magnetic moments near DWs [Fig.~\ref{Fig1}(h)]. 

The model quantitatively reproduces the observed data, both at zero and finite field [see Figs.~\ref{Fig1}(c) and \ref{Fig2}(b-d)]. The fit of the zero field [Fig.~\ref{Fig1}(c)] data at 1.6~K yields that an average distance between DWs is 30.4(1.0)~l.u.[231(8)~\AA] and the distribution function is reasonably sharp with a FWHM of 6(2) l.u. indicating that the domain size is well defined. The presence of a long diffusive tails in the data helps to extract details on the fine structure of the domain walls: the maximal amplitude of the moment suppression is 58(6)~\%\ and the FWHM of the domain wall is 3.1(2) l.u..

We also fitted the data collected with $B\|[001]$ using the same approach. 
The field dependence of mean, mode (the most likely value) and FWHM of the fitted distribution is summarized in Fig.~\ref{Fig2}(e). In the low-field regime, the mode and mean coincide and remain roughly constant up to $\approx2$~T, while the FWHM continuously grows. This increase reflects a continuous broadening of the PDF and corresponds to the suppression of higher-order harmonics at 3$\mathbf{k}_{\rm m}$ and 5$\mathbf{k}_{\rm m}$. Above 2.2~T, the mode rapidly decreases, while the mean, showing a moderate increase at intermediate fields, eventually returns to the initial low-field value. The fitted PDF for three representative fields are summarized in Fig.~\ref{Fig2}(f). One can see a sharp peak at 0.2~T, which broadens considerably at 2~T yet remains at the same position. At higher fields, the peak shifts towards zero reflecting the formation of the short-range ordered state with a finite correlation length determined by the value of the mean.

\textbf{Theoretical model}.
To rationalize the field-induced evolution of the magnetic state,  we consider a model in which nearly isotropic Fe spins are described by the N{\'e}el vector $\mathbf{n} = (\mathrm{cos}(\phi), \mathrm{sin}(\phi), 0)$. For $H\|c$ $\mathbf{n}$ stays in the $ab$-plane, and the AFM ordering of strongly anisotropic Dy spins is described by the Ising variable $\sigma = \pm1$. The free energy density of the IC state in zero fields is:
\begin{align}
    f = \frac{J}{2}\Big(\frac{d\phi}{dz}\Big)^2 + U_{\rm a}(\phi) + \sum_n \Big( f_{\rm Dy}^{(0)} - 2gQ_nn_x(z)\Big)\delta(z-z_n),
    \label{Eq::free_en}
\end{align}
where the first two terms are the exchange and anisotropy of the Fe subsystem. $f_{\rm Dy}^{(0)}$ is the bare free energy of the Dy DW with a topological charge 
$Q_n = \frac{1}{2}[\sigma(z_n+0)-\sigma(z_n-0)] = (-1)^n$ 
located at $z = z_n$. The last term results from the Lifshitz invariant,  $\Gamma_4 \overleftrightarrow{\partial_z} \Gamma_5$, describing the interaction between Fe and Dy moments at DWs. This interaction stabilizes an IC modulation along the $c$-axis, in contrast to TbFeO$_3$, where the $\Gamma_8 \overleftrightarrow{\partial_y} \Gamma_2$ invariant produced an IC modulation along the $b$-axis, $\mathbf{k}_{\rm m} = (0,\pm\delta,1)$. 
Details of calculations and parameter choice can be found in Sec.~S5 of the SM~\cite{SM}.

The stabilization of the IC state has a transparent physical interpretation in the weak coupling limit, when the deviation of $\phi$ from 90$^{\circ}$ is relatively small. In this case, $\phi$ plays the role of a massive scalar field mediating interactions between Dy DWs similar to the Yukawa interaction between nucleons resulting from pion exchange~\cite{kemmer1938nature}. The interaction between two patches of DWs with topological charges $Q_1$ and $Q_2$, separated by a distance $r$, is the screened Coulomb interaction (with opposite sign) $\propto\ -Q_1Q_2\frac{1}{r}e^{-r/l}$, where the screening length $l$ is inversely proportional to the gap in the magnon spectrum  propagating through the Fe magnetic subsystem. The ``dressing'' of the Dy DW by distortions of the Fe ordering decreases its energy:
\begin{align}
    f_{\rm Dy} = f^{(0)}_{\rm Dy}\Bigg(1-\frac{g^2}{f^{(0)}_{\rm Dy}\sqrt{JK_x}}\Bigg)
    \label{eq::Dy_dressing}
\end{align}
where $K_x$ is the strength of the $2^{\rm nd}$-order magnetic anisotropy. Above a critical value of the coupling constant $g$, the DW free energy becomes negative and the IC state appears. Neighboring DWs have opposite topological charges and repel each other, which keeps them at a distance $\propto l$ and determines the IC parameter $\delta$. The IC state suppresses the uniform FE state, which explains why electric polarization was not observed at low magnetic fields~\cite{tokunaga2008magnetic}.

What makes DyFeO$_3$ different from TbFeO$_3$, is the electric and magnetic moments induced by the coexisting Fe and Dy orders:
\begin{equation}
\left\{
\begin{array}{ccc}
P_z &=& \sigma P_0 \cos \phi + \alpha \sigma H_z, \\
M_z&=& M_{WFM} \cos \phi + \alpha \sigma E_z,
\end{array}
\right.
\label{eq:PM}
\end{equation}
where $P_0 = 1.5\cdot 10^3$ $\frac{\rm \mu C}{{\rm m}^2}$ is the spontaneous electric polarization, $M_{wFM} = 0.13$ $\mu_{\rm B}$ is the weak FM moment per Fe ion induced along the $c$-axis by $n_x = \cos \phi$ and $\alpha$ is the linear magnetoelectric coefficient in the uniform Dy state with a $\Gamma_5$ symmetry. Due to depolarization and demagnetization effects, the electric and magnetic fields, $E_z$ and $H_z$, in Eq.~\eqref{eq:PM} differ from the applied fields, $E_0$ and $H_0$. Whereas the weak demagnetization field can be neglected, the depolarization field $\sim 70$ ${\rm kV}\cdot{\rm cm}^{-1}$ effectively modifies the anisotropy of Fe spins and their weak FM moment in the IC state  (see Sec.~S5c in SM~\cite{SM}).

Figures~\ref{Fig3}(c) and (d) show the $z$-dependence of $\phi$  and $\sigma$  in the IC state at $H_0 = E_0 = 0$, in which Dy DWs form an equidistant array and $\phi$ describes the direction of the N\'eel vector oscillating around the average value of $90^\circ$ (dashed line). Figure~\ref{Fig3}(e) shows the corresponding variation of the electric polarization. Negatively charged DWs (the electric charge density, $\rho = - \frac{dP_z}{dz}$) are surrounded by a cloud of compensating positive charges. For $H_0 \neq 0, E_0 = 0$, as well as for $H_0 = 0, E_0 \neq 0$, the DW array remains equidistant and has a zero net polarization. The alternation of the sign of $\sigma$ suppresses the linear magnetoelectric response. However, when both $H_0$ and $E_0$ are nonzero, the array is dimerized [see Fig.~\ref{Fig3} (f, g)] with $P$ parallel (antiparallel) to the applied electric field in longer (shorter) $\sigma$-domains. The dimerization results from the linear magnetoelectric effect.

Figure~\ref{Fig3}(a) shows the DyFeO$_3$ phase diagram in electric and magnetic fields applied along the $c$-axis, obtained by numerical minimization of the free energy using parameters taken from the experiment. Under $H\|c$, the weak FM moment induced by $n\|a$ forces the N\'eel vector in the IC state to tilt (on average) towards the $a$-axis. The decrease of the average angle, $\phi_*$, [dashed line in Fig.~\ref{Fig3}(c)] effectively reduces the Fe-Dy coupling, $g \rightarrow g \sin \phi_*$, which increases the energy of the IC state. At a critical field, $H_{\rm c1}$, the IC state (blue color) undergoes a 1$^{\rm st}$-order transition into  the uniform FE$_1$ state, (gray), in which both Dy and Fe order parameters are uniform, but the N\'eel vector is not yet fully rotated to the $a$ axis. In the FE$_2$ state (red), $\phi = 0$. The transition between the FE$_1$ and FE$_2$ states is abrupt for a sufficiently strong $4^{\rm th}$-order anisotropy: $U_a(\phi) = \frac{K_x}{2}\cos^2\phi + \frac{K_x'}{4}\cos^4\phi$, with $K_x' < 0$. 

The calculated magnetic field dependence of the electric polarization, $P(H)$, shows a very good quantitative agreement with the experimental data by Tokunaga \textit{et al.}~\cite{tokunaga2008magnetic}. 
However, while our theory predicts the formation of a uniform FE state above $H_{\rm c1}$, the experimentally observed order exhibits a large, but finite correlation length [250(60)~\AA\ at 4~T, see Fig.~\ref{Fig2}(b)], related to the metastability of the DW array and pinning of DWs by disorder. The DW pinning can further explain the minor difference between the propagation wavevector in our samples and~\cite{ritter2022magnetic}. 

The behavior of $H\|b$ can be understood based on the fact that DWs in the Ising-like $A_y$-ordering of Dy spins carry a FM moment (anti)parallel to the $b$-axis. A small adjustment of the DW positions therefore aligns the magnetic moments of all DWs along the magnetic field direction, which reduces the DW free energy: $f^{(0)}_{\rm Dy} \rightarrow f^{(0)}_{\rm Dy} - \cos \alpha_{\rm Dy} \mu_{\rm Dy} H_b$, where $\mu_{\rm Dy} = 10 \mu_{\rm B}$ is the Dy magnetic moment and $\alpha_{\rm Dy}$ is the angle between the spin-quantization axis of Dy ions and the $b$-axis. The result is the decrease of the period of the IC state for $H\|b$, until the uniform $F_y C_x$-state sets in (see Fig.~\ref{Fig_010}).

\textbf{Discussion and conclusion.}
Frustration due to interactions between rare-earth and Fe spins in orthoferrites can be lifted in two different ways: by polar lattice distortions inducing an electric polarization or by periodic modulations of the two magnetic orders, which gives rise to the competition between uniform multiferroic and IC states. 
In this work we used a combination of high-resolution neutron diffraction, Monte-Carlo simulations, and model calculations 
to understand the complex low-temperature behavior of \dfo.
We showed that in low fields DWs in the AFM ordering of Dy spins form a periodic array (soliton lattice) stabilized by the interactions with Fe spins at the DWs.
Due to long-ranged interactions mediated by magnons propagating through the Fe magnetic sublattice, the sharp Dy DWs effectively acquire a large width, comparable to the width of domain walls in the AFM ordering of Fe spins, which explains the long period of the IC state. 
The IC state is the ground state of the system, but it can be suppressed by the application of magnetic and/or electric fields. 
A magnetic field applied within the $ab$-plane couples directly to Dy moments, reduces the domain size and eventually stabilizes field-induced FM state. 
By contrast, the interaction of the $c$-axis magnetic field with Dy spins parallel to the $ab$ plane is weak.
Instead, it causes the re-orientation of Fe spins toward the $a$-axis, which suppresses the IC state and induces a uniform AFM order of Dy spins with a finite correlation length, presumably due to remaining randomly positioned DWs.  
This state has a spontaneous electric polarization due to polar shifts of Dy ions and shows a giant linear magnetoelectric effect observed in Ref.~\cite{tokunaga2008magnetic}.

The linear magnetoelectric coupling allows to control the magnetic soliton lattice: When both electric and magnetic fields are applied along the $c$-axis, the IC state dimerizes -- the widths of neighboring domains become unequal [Fig.~\ref{Fig3}(d,g)]. 
We estimate that in zero field the IC state is stable up to high electric fields of $\sim350$~kV/cm. 
However, when a magnetic field is applied, the critical electric field is strongly reduced and comes down to the field range used in the study of electric polarization of \dfo{}~\cite{tokunaga2008magnetic}: $E = 80$~kV/cm at $\mu_0H_0 = 1$~T and $E = 35$~kV/cm at $\mu_0H_0 = 1.5$~T [see Fig.~\ref{Fig3} (a)]. 
It is still a question whether the Dy DWs can be moved and annihilated at low temperatures. 
However, our calculations show, that the electric control of magnetism in \dfo\ might be possible close to  $H_{\rm c1}$.

Our analysis indicates that the ground state of \dfo\ is determined by a delicate balance between spin interactions and magnetic anisotropies, and can be tuned between the magnetic soliton lattice and uniform ferroelectric state by the application of magnetic and/or electric fields. This result is important for electric control of magnetism in a broad class of multiferroic materials with several coexistent magnetic orders.

\section{Methods}
\noindent \textbf{Sample preparation.}

The single crystal of \dfo\ was grown by the floating-zone method using an optical floating-zone furnace (FZ-T-10000-H-IV-VP-PC, Crystal System Corp., Japan) with four 1000~W halogen lamps as a heat source as described in~\cite{biswas2022role}. \dfo\ crystallizes in $Pbnm$ space group with low-temperature lattice parameters, $a = 5.293$~\AA, $b = 5.584$~\AA\ and $c = 7.587~$\AA. 
For the neutron scattering experiments we used a 
\dfo crystal with a mass of $m = 19$~mg~\cite{biswas2022role}.

\noindent \textbf{Neutron scattering experiments.}

We performed three neutron diffraction measurements using the cold-neutron Triple-Axis Spectrometer (TASP) at SINQ, PSI. In the first experiment, the sample was oriented in the $(h,0,l)$ scattering plane and installed in the orange cryostat. For the first in-field measurements, we oriented the crystal in the $(h,h,l)$ scattering plane and a horizontal cryomagnet MA-7 was used to produce magnetic field. In the second in-field experiment, we oriented the sample in the $(h,0,l)$ plane and applied a vertical field along the $b$-axis using the MA-10 cryomagnet. The base temperature in all experiments was $\approx 1.7$~K.
We operated the instrument in the diffraction mode (no analyzer used) with $k_{\rm i} = 1.3$~\AA$^{-1}$. Two 40' collimators were installed before and after the sample. A cold Be filter was used to suppress a higher order harmonic contribution.

\noindent \textbf{Monte-Carlo fits of the diffraction data.}

To simulate the observed diffraction profile, we considered a one-dimensional Ising chain with domain wall-type defects $...\!\uparrow\downarrow\uparrow\uparrow\downarrow\uparrow\!...$. The chain size was up to $N = 2^{19} = 524288$.  The distance between the domain walls was assumed to follow a log-norm distribution. 
To describe the effect of Dy DW ``dressing'' by the Fe subsystem, we assume that the DWs within the Dy subsystem have a finite size. To achieve this, the amplitude of the Dy moments near the domain walls was suppressed by a Gaussian profile as schematically shown in Fig.~\ref{Fig1}(h) and described in details in Sec.~S4 of the SM~\cite{SM}. The diffraction intensity was calculated using the standard Fast Fourier Transform. 
All details of the calculations can be found in Sec.~S4 of SM~\cite{SM}.

\noindent \textbf{Theory of IC state in \dfo.}

The $z$-dependence of the Fe and Dy order parameters in the IC state and the phase diagram were obtained by a numerical minimization of the free energy   Eq.~\eqref{Eq::free_en}, to which we added the coupling to applied electric and magnetic fields, as well as the depolarization and demagnetization energies. All details regarding the model and choice of parameters can be found in Sec.~S5 of SM~\cite{SM}, where we also discuss the weak coupling approximation that works well for DyFeO$_3$.

\bibliography{bibliography}

\smallskip
\noindent\textbf{Data availability}
All relevant data are available from the authors upon reasonable request.

\noindent\textbf{Competing interests}
The authors declare no competing interests.

\noindent\textbf{Acknowledgments}
 We acknowledge stimulating discussions with Dr. Tom Fennell. 
 We acknowledge financial support from the Swiss National Science Foundation (Project No.~200020-169393), the German Research Foundation (DFG) through the Collaborative Research Center SFB 1143 (project \# 247310070); through the W\"urzburg-Dresden Cluster of Excellence on Complexity and Topology in Quantum Materials\,---\,\textit{ct.qmat} (EXC~2147, Project No.\ 390858490). This work is based on experiments performed at the Swiss spallation neutron source SINQ, Paul Scherrer Institute, Villigen, Switzerland. \smallskip

\end{document}


\title{Supplemental Material\\ Competition Between Multiferroic and Magnetic Soliton Lattice States in \dfo}

\maketitle



\section{Results of PCM calculations}\label{smsec::2}
We performed PCM calculations using the \textsc{McPhase} software~\cite{mcphase}. The calculated $B_l^m$ parameters are summarized in Table~\ref{tab:blm} (here we report the CEF parameters using in the coordinate system with $x,y,z$ parallel to the $a,b$ and $c$ axes of the crystal. To verify the accuracy of our PCM model, we calculated the saturation magnetization along the three crystal axes and found $M_a = 4.8~\mu_{\rm B}, M_b = 8.8~\mu_{\rm B}$ and $M_c < 0.3~\mu_{\rm B}$. These results are in excellent quantitative agreement with the experimental data from~\cite{zhao2014ground}  taking into account the simplicity of the model ($M_a = 4.4~\mu_{\rm B}, M_b = 8.6~\mu_{\rm B}$ and $M_c = 0.9~\mu_{\rm B}$ at $T = 2$~K and $B = 7$~T). The difference in the $c$-axis signal likely arises from the weak FM component of the Fe subsystems, not included in the calculation.

To calculate the ground state wavefunctions we rotate the quantization axis along the easy axis of the Dy moments as schematically shown in Fig.~1(e) of the main text. 
In this case the wavefunctions of the groundstate the and first excited doublets (which lies at 28.01~meV) are given by:
\begin{align}
|\psi_{\pm}\rangle_0 = 0.997 |\pm15/2\rangle +... \\
|\psi_{\pm}\rangle_1 = 0.999 |\pm13/2\rangle +... 
\end{align}
Both wavefunctions constitute of almost pure $|\pm15/2\rangle$ and $|\pm13/2\rangle$  states respectively having only negligible admixture of other states. Thus, we conclude that Dy magnetism at low temperatures, $<$10~K, can be well described by the pseudo-$S=1/2$ model with an extreme Ising-like anisotropy.

\begin{table}[b!]
\label{tab:blm}
\renewcommand{\arraystretch}{1.25}  
\small
\caption{~The set of $B_l^m$ parameters in units (meV) calculated using PCM.}
\begin{tabular}{llcllc}

	\hline\hline
	
	$B_l^m$ (meV)&   &   &   &   &   \\
	
	\hline
	
	$B_2^0$ & = & $4.30\times10^{-1}$ & $B_6^0$ & = & $-6.24\times10^{-7}$\\
	
	$B_2^2$ & = & $2.63\times10^{-1}$ & $B_6^2$ & = & $2.26\times10^{-6}$\\
	
	$B_2^{-2}$ & = & $-5.53\times10^{-1}$ & $B_6^{-2}$ & = & $2.68\times10^{-6}$ \\
	
	$B_4^0$ & = & $4.79\times10^{-5}$ & $B_6^4$ & = & $-1.53\times10^{-5}$\\
	
	$B_4^2$ & = & $5.64\times10^{-4}$ & $B_6^{-4}$ & = &  $-6.77\times10^{-7}$\\
	
	$B_4^{-2}$ & = & $-1.43\times10^{-3}$ & $B_6^6$ & = & $2.15\times10^{-6}$ \\
	
	$B_4^4$ & = & $4.61\times10^{-4}$ & $B_6^{-6}$ & = & $6.21\times10^{-7}$\\
	
	$B_4^{-4}$ & = & $4.01\times10^{-3}$&  &  &  \\
	\hline\hline
	
\end{tabular}
\end{table}

\section{Neutron diffraction in [110] fields}

\begin{figure}[tb]
\center{\includegraphics[width=1\linewidth]{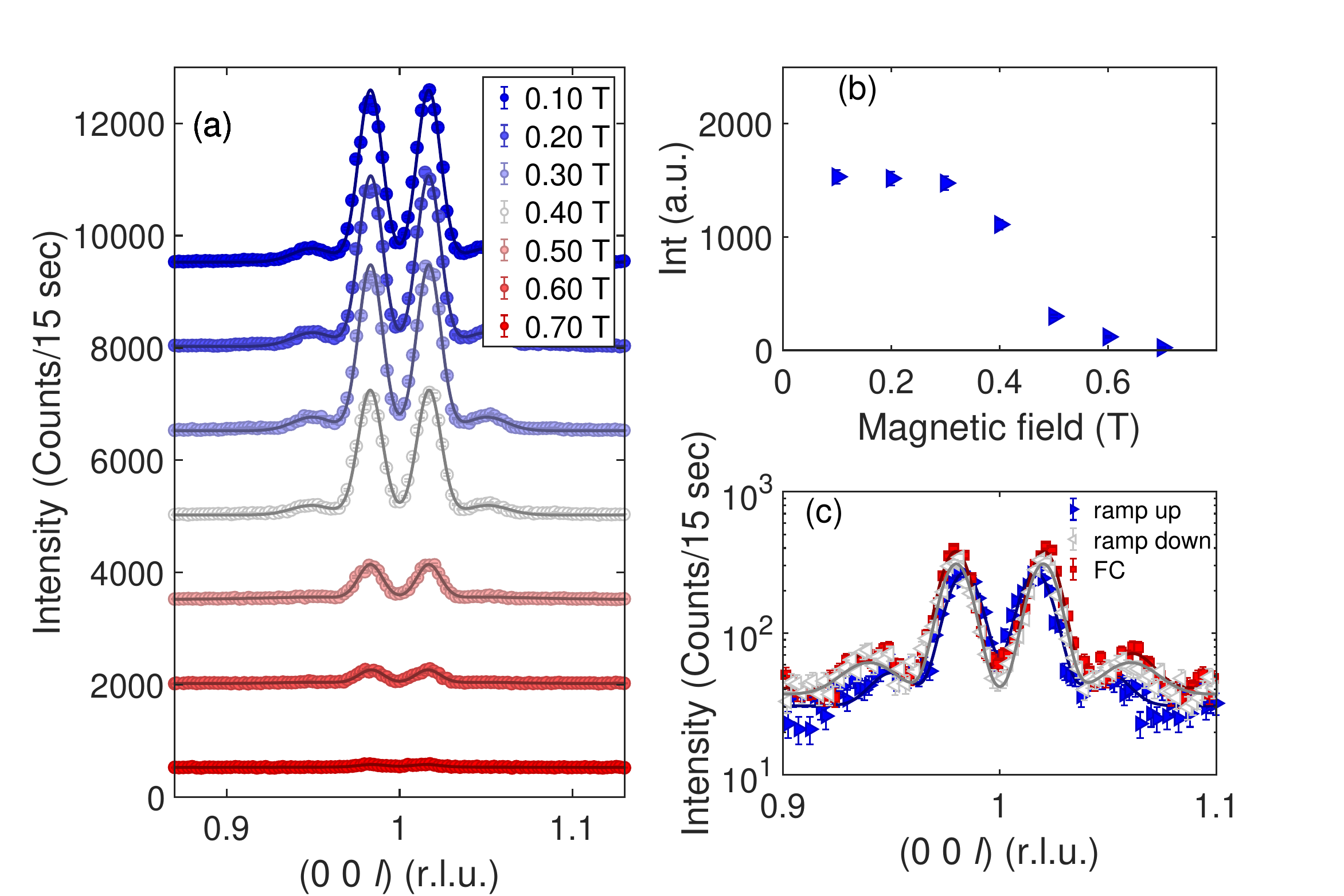}}
  \caption{~(a)~Neutron diffraction scans measured along $(0,0,l)$ directions at $T = 2$~K and various fields applied along $[1,1,0]$ the direction as indicated in the legend.  Data are stacked by 1500 units for clarity
  (b)~Intensity of $(0,0,1\pm\delta)$ peaks extracted from the data in panel (a).
  (c)~Scans along $(0,0,l)$ direction taken at $T = 2$~K and $B = 0.6$~T state, prepared using different protocols. Blue curve was collected after ZFC and then sweeping the field to 0.6~T.
  Grey curve was collected after FC at 0.6~T. 
  Red curve was measured after sweeping the field down to 0.6~T from 1.6~T.
  }
  \label{Fig_SM_field_110}
  \vspace{-12pt}
\end{figure}

Additional measurements were performed with the magnetic field applied along the [110] direction. Figure~\ref{Fig_SM_field_110}(a) shows the scans taken at $T = 2$~K and various fields, $B\|[110]$ applied using the horizontal MA-7 cryomagnet. One can see that the behavior changes drastically as compared to the $B \| [001]$ case. The magnetic field rapidly suppresses the magnetic satellites, which become undetectable already at 0.7~T [Fig.~\ref{Fig_SM_field_110}(b)], however the propagation wavevector remains unchanged. To see whether the magnetic prehistory influences the observed signal, we performed scans at 0.6~T and 2~K using different cooling procedures: (i) ZFC and then apply 0.6~K field; (ii) FC at 0.6~T; (iii) FC in 1.6~T and then decrease the field to 0.6~T. The results are summarized in Fig.~\ref{Fig_SM_field_110}(c). Our data show a small change in intensity of the satellites indicating a hysteretic behavior in agreement with~\cite{wang2016simultaneous}. Moreover, the propagation wavevector also exhibits a moderate change from 0.0165(2) (ZFC) to 0.0201(2) (FC).

\section{Empirical fits of the diffraction data}
\begin{figure}[tb!]
\center{\includegraphics[width=1\linewidth]{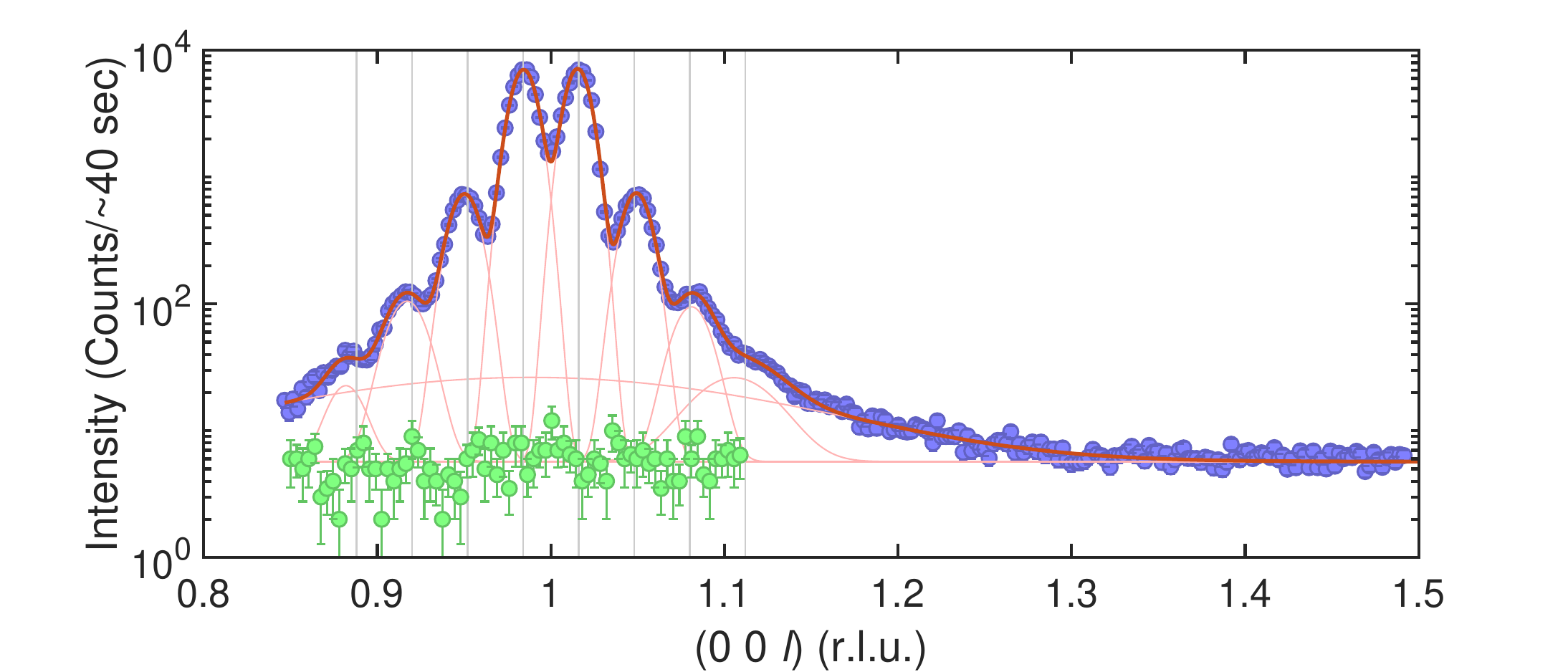}}
  \caption{~Neutron diffraction scans measured along the $(0,0,l)$ directions at $T = 1.7$ and 4.5~K. The data were corrected for the absorption.
  The data are fitted as a sum of 9 Gaussian functions describing 1st, 3rd, 5th, 7th harmonics and a broad peak responsible for the long tail at $l > 1.1$. Blue and green dots show data taken at 1.7 and 4.5~K, the dark red line is the total fit and the light red curves demonstrate the decomposition of the total fit to individual Gaussians. 
  The incommensurabillity parameter is best determined for the first two peaks at $(0,0,1\pm\delta)$. Vertical grey lines show the calculated positions of the higher harmonics based on the $\delta$ value obtained for the first harmonics.
  }
  \label{Fig_SM_fits}
  \vspace{-12pt}
\end{figure}

\begin{figure}[tb!]
\center{\includegraphics[width=1\linewidth]{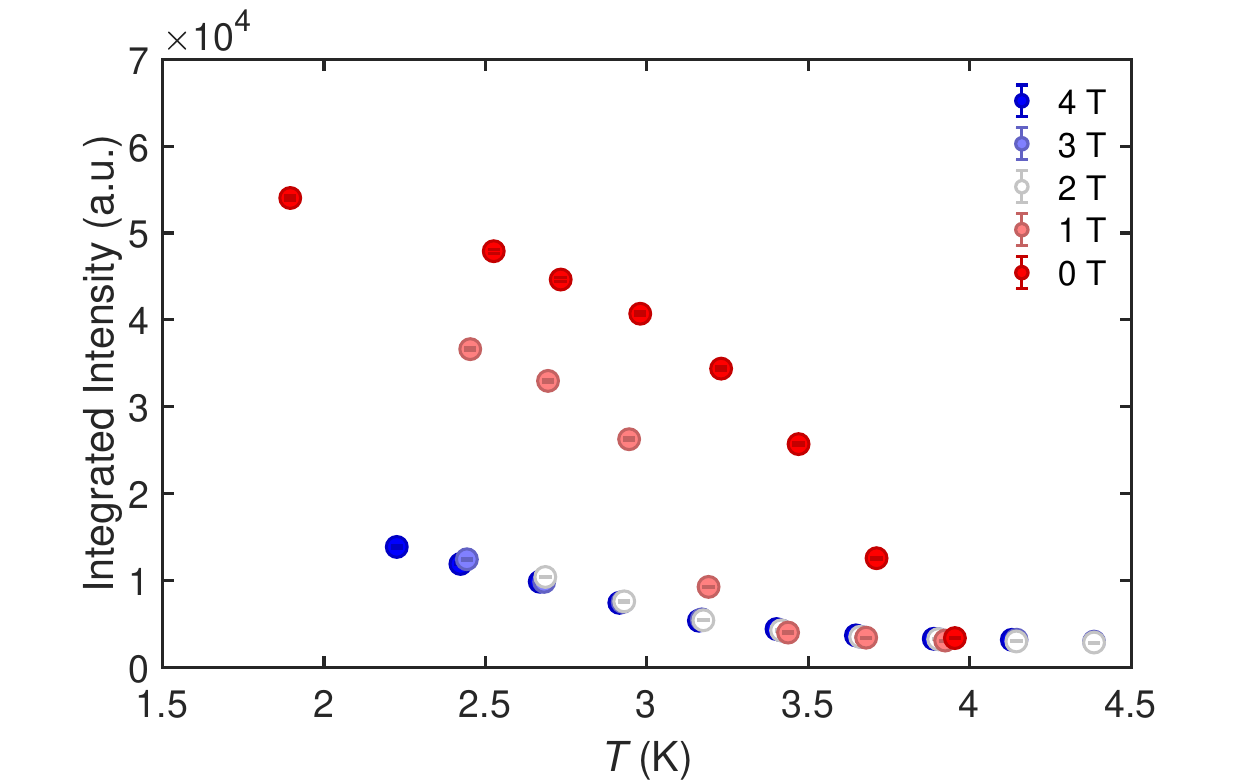}}
  \caption{~Temperature dependence of integrated magnetic intensity measured at different fields applied along $c$ axis.
  }
  \label{Fig_SM_T_dep_AFM}
  \vspace{-12pt}
\end{figure}

In Fig.~1(f,g) of the main text, we show temperature dependencies of the intensity and position of the primary AFM peaks, $(0,0,1\pm\delta)$. They were obtained by empirical fits to the data using a sum of Gaussian functions peaked at odd harmonics, $n = 1,3,5$ and 7. The broad tail at $l > 1.1$ were empirically described by a broad Gaussian centered at $l = 1$. An example of such fit for the base temperature is shown in Fig.~\ref{Fig_SM_fits}. 

Figure~\ref{Fig_SM_T_dep_AFM} demonstrates the temperature dependence of the integrated magnetic intensity measured at different magnetic fields applied along the $c$-axis.

\section{Details of Monte-Carlo calculations}

To simulate the observed diffraction profile, we considered a 1D Ising chain, the diffraction pattern of which was simulated with the help of the Monte-Carlo method. The standard Fast Fourier Transform (FFT) was employed to calculate the diffraction pattern of the sampled chain. Therefore, it was convenient to set the chain size $N$ as a power of two, 2$^n$, in order to take advantage of the FFT algorithm. In our case, $N = 2^{19} = 524288$ the number of spins was chosen and deemed sufficiently large. The calculated Fourier transform was then scaled using the magnetic atomic form factor of Dy$^{3+}$ to better approximate the experimental neutron scattering intensity. Additionally, an absorption correction was applied to the calculated diffraction intensity, to account for strong absorption of Dy nuclei. The absorption was estimated using the \textsc{DAVE} software~\cite{Azuah_2009}, assuming the cylindrical shape of the sample. Finally, to obtain the calculated intensity, the corrected Fourier transform was convoluted with the experimental resolution function taken as a Gaussian with $\sigma = 0.0073$~r.l.u. (calculated using \textsc{TAKIN}~\cite{weber2016takin} software).

We considered an AFM spin chain with domain wall-type defects in a manner of $...\!\uparrow\downarrow\uparrow\uparrow\downarrow\uparrow\!...$. In order to conveniently compare the results with \dfo, the unit cell of the chain was selected to include two spins. Because of that, in the absence of defects, the main reflection occurs at $L = 1$, corresponding to $\mathbf{Q} = (0, 0, 1)$ in \dfo. Once the periodic domain walls are introduced, the peak splits into IC satellites at $L = 1\pm\delta$ along with corresponding higher order harmonics, as discussed in the main text. The distance between the defects (i.e. the size of the AFM domains) directly controls the $\delta$ parameter. However, in addition to the IC parameter, other features appear in the experimental pattern: the intensities of the higher harmonics, the intrinsic width of the peaks, and the tails at higher $Q$ values. These features are governed by other properties of the Ising spin chain, such as the statistical characteristics of the defect distribution, which may exhibit varying degrees of disorder. Different assumptions about the defect configurations can be made, but we first discuss the simplest case.

\begin{figure}[tb]
\center{\includegraphics[width=1\linewidth]{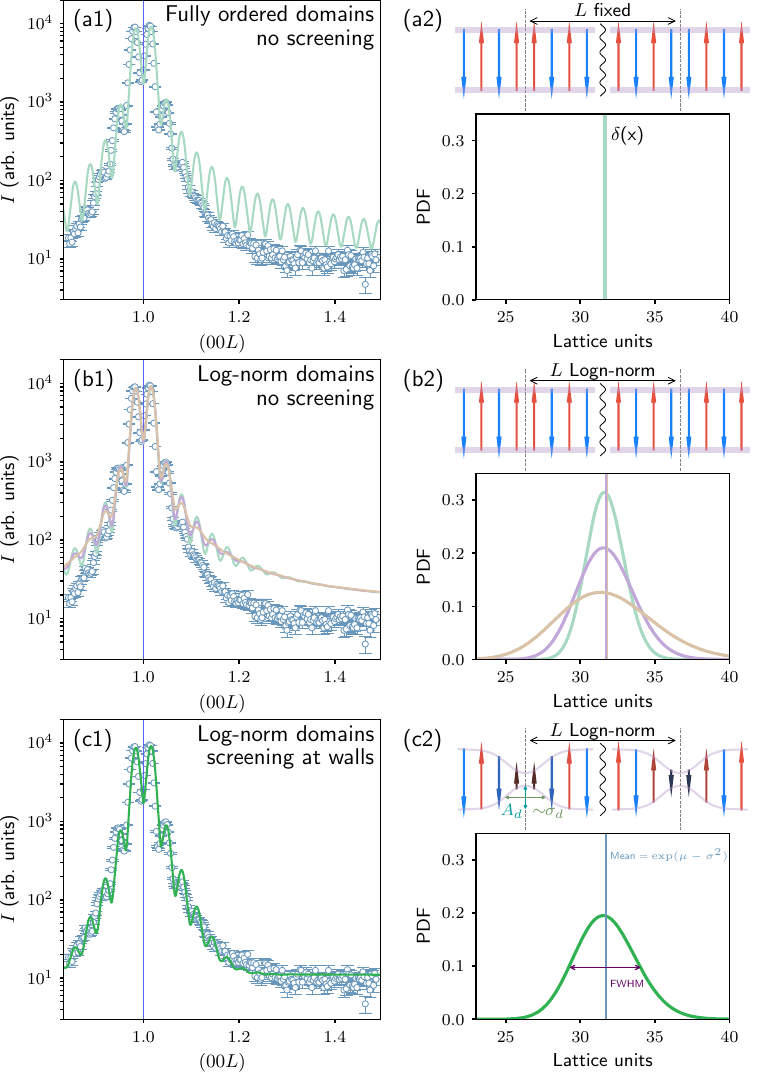}}
  \caption{~Monte-Carlo simulations. The left-side panels are diffraction patterns: the blue points are neutron diffraction at 1.5~K in zero field, the solid lines are the simulated patterns. The right-side panels show the type of used domain walls and probability density function (PDF). (a1,a2)~The case of fully ordered domains with abrupt domain walls. (b1,b2)~The partially disordered defects, abrupt domain walls. (c1,c2)~The partially disordered defects, ``screened'' domain walls.
  }
  \label{FigS4}
  \vspace{-12pt}
\end{figure}

If we assume a fully ordered domain wall pattern with a fixed distance between the defects, the resulting diffraction pattern is very distinctive [Fig.~\ref{FigS4}(a)]. The higher harmonics show no intrinsic width, with only instrumental broadening being present. The intensity of the higher-order satellites decays much slower compared to the experiment. These discrepancies arise due to the partial disorder in the domain configuration of \dfo.

To address this, the next step is to move from a fully ordered configuration towards a Monte-Carlo approach. Here, we assume that the defects are separated by random distances that follow a certain probability distribution. Given the partial disorder, a large system with numerous domains is required to produce a consistent and well-controlled diffraction pattern. Since there is no a prior knowledge of the statistical distribution of the domain configurations we adopt a log-norm distribution, which is a natural choice for positively defined variables such as time intervals, sizes or distances. The probability density function (PDF) for a log-norm distribution can be written as follows:

\begin{align}
    P(x) = \frac{1}{x\sigma\sqrt{2\pi}}\exp{\bigg(-\frac{(\ln{x} - \mu)^2}{2\sigma^2}\bigg)},
    \label{Eq:lognorm}
\end{align}
where parameters $\sigma$ and $\mu$ are the logarithm of scale and the logarithm of location, respectively. Since the distribution is not symmetrical, the mode (the most probable value) and the mean are not the same and are defined as $\exp(\mu - \sigma^2)$ and as $\exp(\mu + \sigma^2/2)$ respectively. The log-normal FWHM is given by: $\exp[(\mu - \sigma^2) + \sqrt{2\sigma^2\ln{2}}] - \exp[(\mu - \sigma^2) - \sqrt{2\sigma^2\ln{2}}]$.

Using this approach, we can sample large chains with domain configurations where the randomness is controlled by the parameters of the log-normal probability density function (PDF). Examples of the diffraction patterns obtained for three representative sets of PDF parameters are shown in Fig.~\ref{FigS4}(b). The introduction of partial disorder leads to a broadening of the higher-order harmonics, bringing the simulated pattern much closer to the experimental data. However, the intensity in the experimental data decays much faster with increasing $Q$. The model overestimates the intensity of the tails, and we found that this discrepancy cannot be resolved by adjusting the PDF parameters. While varying $\sigma$ affects the smearing of the peaks, it does not change the rate of intensity decay. Furthermore, the parameter $\mu$ is constrained by the propagation vector, which must align with the mean of the distribution. We also considered several other PDF functions (Gaussian, Lorentzian, hat, box), but it still did not allow us to reduce the high-$Q$ intensity. Therefore, additional degrees of freedom are needed in the model to accurately reproduce the diffraction pattern of \dfo.

Instead of {\it sharp}, abrupt defects, the domain walls are often smooth and have a finite size. In the presence of extreme anisotropy, the Ising spins have only one degree of freedom\,---\,the magnitude of the magnetic moment\,---\,which can continuously change across the defect. In this scenario, the magnetic moments have a reduced magnitude at the defect location, gradually regaining their nominal value deeper within the domain [Fig.~\ref{FigS4}(c2]. Physically, the origin of the reduction of Dy moment at the domain wall can be related to the interaction between Dy and the Fe subsystems, because the Fe subsystem is polarized by Dy domain walls and provides an effective ``screening'' of magnetic moment.

The screening can be included into the model by multiplying the magnetic moments by a reduction factor, which follows a Gaussain function taking the distance from the domain wall as argument. This Gaussian screening has two parameters: An amplitude $A_s$, which dictates how strong the screening is, and $\sigma_{s}$, which defines the extent of the screening (or, equivalently, the domain wall size). The top of Fig.~\ref{FigS4}(c2) illustrate how the screening compares to the {\it sharp} domain walls [Fig.~\ref{FigS4}(a2,b2)].

The final Monte-Carlo diffraction pattern with refined parameters shown in Fig.~\ref{FigS4}(c) demonstrate an excellent agreement with experiment. The parameters of the log-normal distribution of domain wall distances are deduced from the broadening of higher harmonics, while the screened defects yield the correct decay of the intensity tails. The standard errors were obtained by calculating the $\chi^2$ with varying parameters. For the diffraction in zero field at 1.5~K the parameter set was fitted with the following results: $\sigma = 0.065(13)$; $\mu = 3.455(8)$; $A_s = 58(6)\%$; $\sigma_s = 1.33(11)$.

\section{Theory of the IC state in Dysprosium orthoferrite}

\subsection{Weak coupling approximation} 

In the absence of applied fields, the direction of the N\'eel vector is close to the $b$ axis. Expanding free energy in powers of 
$\varphi(z) = \frac{\pi}{2} - \phi(z)$, we obtain
\begin{equation}
\left[-J \frac{d^2}{dz^2} + K_x\right]\varphi(z) = 2g\sum_n Q_n \delta(z-z_n).
\end{equation}
The solution is: 
\begin{equation}
\varphi(z) = \frac{g}{\sqrt{JK_x}}\sum_n Q_n e^{\frac{|z-z_n|}{l}},
\end{equation}
where $l = \sqrt{\frac{J}{K_x}}$. 

Substituting $\varphi(z)$ into Eq.(1) of the main text, we obtain free energy of the domain wall array:
\begin{align}
F - F(\varphi=0) = 
N f_{\rm Dy}^{(0)}
-  \frac{g^2}{\sqrt{JK_x}} 
\sum_{n,m} Q_n Q_m e^{\frac{|z_n-z_m|}{l}},
\end{align}
where $F(\varphi=0)$ is the free energy of the uniform state with $n\|b$ and $N$ is the number of domain walls. The terms with $n = m$ in the double sum over domain wall positions describe the self-interaction that lowers the domain wall free energy: $f_{\rm Dy}^{(0)} \rightarrow f_{\rm Dy} = f_{\rm Dy}^{(0)}(1-\lambda)$, $\lambda = \frac{g^2}{f_{\rm Dy}^{(0)}\sqrt{JK_x}}$ being the dimensionless coupling constant. The interaction between two domain walls (per unit area of the wall) is
\begin{equation}
U(z_n-z_m) =  - \frac{2 g^2}{\sqrt{JK_x}}  Q_n Q_m e^{\frac{|z_n-z_m|}{l}}.
\end{equation}  
This result can also be obtained by integrating ``the screened Coulomb potential'', $- \frac{g^2}{\pi J} Q_n Q_m \frac{1}{r} e^{-r/l}$, over the domain wall surface. The ``screening" originates from the gap in the magnon spectrum. This shows that Dy domain walls behave like particles that interact by emitting and absorbing virtual magnons propagating through the Fe magnetic sublattice.  

The free energy of an equidistant domain wall array with $z_n = n d$ is given by
\begin{equation}
F - F(\varphi=0) = N f_{\rm Dy}^{(0)}
	\left(1 -\lambda \tanh\frac{d}{2l}\right),
\end{equation}
and the distance $d$ between the domain walls is found by minimizing the free energy of the domain wall array per unit length. 

\subsection{Coupling to electromagnetic field}

The interaction with the electric field, $E_0$, and magnetic field, $H_0$, applied along the $c$ axis is described by
\begin{align}
f_{\rm int} = - E_0 P_z - H_0 M_z + \frac{2\pi}{\epsilon} \left(P_z - \bar{P_z}\right)^2 + 2 \pi M_z^2,
\label{eq:fint}
\end{align}
where the electric polarization, $P_z$, and magnetization, $M_z$, are given by Eq.(3) in the main text. The last terms are, respectively, the depolarization and demagnetization energies (here, we use the CGS system of units). We assumed, for simplicity, that the sample has the shape of a thin film of thickness $L_z$ in the $c$ direction and that the electric field is applied by ideal electrodes attached to its upper and lower surfaces normal to the $c$ axis. The electric field, $E_z$, in Eq.(3) is related to the applied electric field, $E_0 = \frac{U}{L_z}$, $U$ being the applied voltage, by
\begin{equation}
E_z = E_0 + \frac{4\pi}{\epsilon}\left(\bar{P_z} - P_z\right),
\end{equation}
where $\bar{P_z}$ is the average polarization and $\epsilon = 23$ is the background electric permittivity~\cite{stanislavchuk2016magnon}, which follows from the fact that the $z$-component of the displacement field, $D_z = \epsilon E_z + 4\pi P_z$, is constant and equal to its average value, and from $\bar{E_z} = E_0$. In the uniform state with $\bar{P_z} = P_z$ the depolarization energy is zero, because the surface charges are screened by electrodes. On the other hand, the depolarization energy is nonzero in the domain wall array state due to the electric charges induced in the bulk. The close competition between the IC and ferroelectric states, and the large depolarization field value, $E_{\rm dep} = \frac{4\pi P_0}{\epsilon} \approx 73$ kV/cm, make the depolarization energy important.

 The demagnetization field induced by the weak ferromagnetic moment, $4 \pi M_{\rm WFM} \approx 270$ G, as well as the magnetic moment induced by applied electric fields, are small, so that the demagnetization effects are only important for very weak magnetic fields and can be neglected. There is also a term quadratic in the applied magnetic field, $\frac{\chi_\perp}{2}
 \left[
 \left(\mathbf{n}\cdot \mathbf{H}\right)^2 - H^2
 \right]$, where $\chi_\perp$ is the transverse magnetic susceptibility of the antiferromagnetically ordered state of Fe spins. However, for the magnetic fields used in our experiment, this term is relatively small. Furthermore,  for $H\|c$, $\mathbf{n}$ rotates in the $ab$ plane and is perpendicular to $\mathbf{H}$. The interaction of the Ising-like Dy spins parallel to the $ab$ plane with $H\|c$ can be neglected.

 \subsection{Numerical simulations of the IC states}

 The calculation of the IC state spin configuration in the presence of applied electric and magnetic fields is complicated by the fact that the depolarization energy depends on the average polarization, $\bar{P}_z$, that itself depends on the applied fields, $\phi(z)$ and domain wall coordinates. This problem does not arise if only $E_0$ or $H_0$ is applied, as in these cases $\bar{P}_z = 0$. In the following, we describe the self-consistent calculation of $\phi(z)$ and domain wall positions in applied electromagnetic fields. 

 Minimizing the sum of Eqs. (1) and (\ref{eq:fint}), we obtain the following equation for $\phi(z)$ in the IC state:
 \begin{align}
     - J \frac{d^2 \phi}{dz^2} &- \left(K_x+\frac{4\pi P_0^2}{\epsilon}\right) \cos\phi \sin\phi - K'_x \cos^3\phi \sin\phi \nonumber\\
     &+  P_0  
     \left(E_0 + \frac{4\pi \bar{P}_z}{\epsilon} 
     + \frac{4\pi\alpha}{\epsilon}H_0\bar{\sigma}\right)\sigma \sin\phi
     \nonumber \\
     &+ H_0 \left(M_{\rm WFM} - \frac{8 \pi \alpha P_0} {\epsilon}\right) \sin\phi\nonumber\\ 
     &+ 2 g \sum_n Q_n \sin \phi(z_n) \delta(z-z_n)=0,
 \end{align}
 where $\bar{\sigma}$ is the average value of the Dy order parameter. Here, we took into account that the average polarization, $\bar{P}_z$, is a functional of $\phi(z)$. For numerical simulations, it is more convenient to do minimization at constant $\bar{P}_z$. The corresponding functional that leads to the same equation for $\phi(z)$ has the form,
 \begin{align}
 F' = \int\!\!dz \Bigg[
&\frac{J}{2}\left(\frac{d\phi}{dz}\right)^2 + \frac{1}{2}
\left(K_x+\frac{4\pi P_0^2}{\epsilon}\right)\cos^2 \phi + 
\frac{K'_x}{4}\cos^4\phi \nonumber \\
&-\frac{P_0}{\epsilon}  
     \left(\epsilon E_0 + 4\pi \bar{P}_z 
     + 4\pi\alpha H_0\bar{\sigma}\right) \sigma \cos \phi
     \nonumber\\
&-H_0 \left(M_{\rm WFM} - \frac{8 \pi \alpha P_0} {\epsilon}\right) 
\cos \phi \nonumber\\
&+\sum_n\left(f^{(0)}_{\rm Dy} -2 g \sum_n Q_n \cos \phi \delta(z-z_n)\right)
\Bigg].    
 \end{align}
 The depolarization energy modifies the second-order magnetic anisotropy and the weak ferromagnetic moment. Importantly, $F'$ depends on $E_0$ and 
 $\bar{P}_z$ through the displacement field,  $D_z=\epsilon E_0 + 4\pi \bar{P}_z$. First, we find $\phi(z)$ for a given $D_z$ and positions of the domain walls, calculate  $\bar{P}_z$ and  $\bar{\sigma}$, minimize the free energy density with respect to the domain wall positions and re-plot it as a function of $E_0$. 
 
 \subsection{Parameter choice}
 
 Simulations are performed, for $J = 4.5 S^2$ meV, $S = 5/2$ being the spin of the Fe ion, similar to that found for YFeO$_3$ and other orthoferrites~\cite{Hahn, skorobogatov2020low, Podlesnyak2021}. We chose $f_{\rm Dy}^{(0)} = 2.6$~K (the ordering temperature of Dy spins, $T_{\rm Dy} \sim 4.5$~K). The choice of the Fe-Dy coupling constant $g$ and the anisotropy parameter $K_x$ was made as follows. We found that for $\lambda \gg 1$, the IC state period strongly increases with an applied magnetic field $H\|c$ before this state disappears at $\sim 2$~T, which is not observed in our experiment.  On the other hand, when $\lambda$ is close to the critical value 1, the IC state disappears at very low applied fields. We use $g = 8.5$ K and  $K_x  = 0.5$ K, corresponding to $\lambda = 2.2$, for which the period changes between $58 c$ in zero field and $62 c$ at $H_{\rm c1} = 2.2$~T, when the IC state starts to disappear in our experiment. In calculations, it undergoes a first-order transition into a uniform ferroelectric state with $\phi > 0$ at $H_{\rm c1}$. Tokunaga \textit {et al.}~\cite{tokunaga2008magnetic} found that spontaneous polarization saturates and the magnetic field dependence of $P$ solely results from the linear magnetoelectric effect above $H_{\rm c2} \approx 2.6$~T, corresponding to the disappearance of IC peaks in our experiment, which suggests that at $H_{\rm c2}$ the N\'eel vector becomes parallel to the $a$ axis.  We reproduce two first-order transitions at $H_{\rm c1}$ and $H_{\rm c2}$ by adding the fourth-order anisotropy term with $K'_x  = -0.32$~K. The slope of $P(H)$ above $H_{\rm c2}$ is reproduced using $\alpha = 0.0072$ in dimensionless CGS units ($\alpha = 0.024$ reported in~\cite{tokunaga2008magnetic} is apparently given in $\frac{\mu{\rm C}}{{\rm cm}^2 \cdot {\rm T}}$, which can be deduced from Fig. 3b in that paper). For $\lambda =2.2$, the weak coupling approximation and numerical simulations give close results.

\subsection{Domain wall array dimerization}
\begin{figure}[tb!]
\center{\includegraphics[width=1\linewidth]{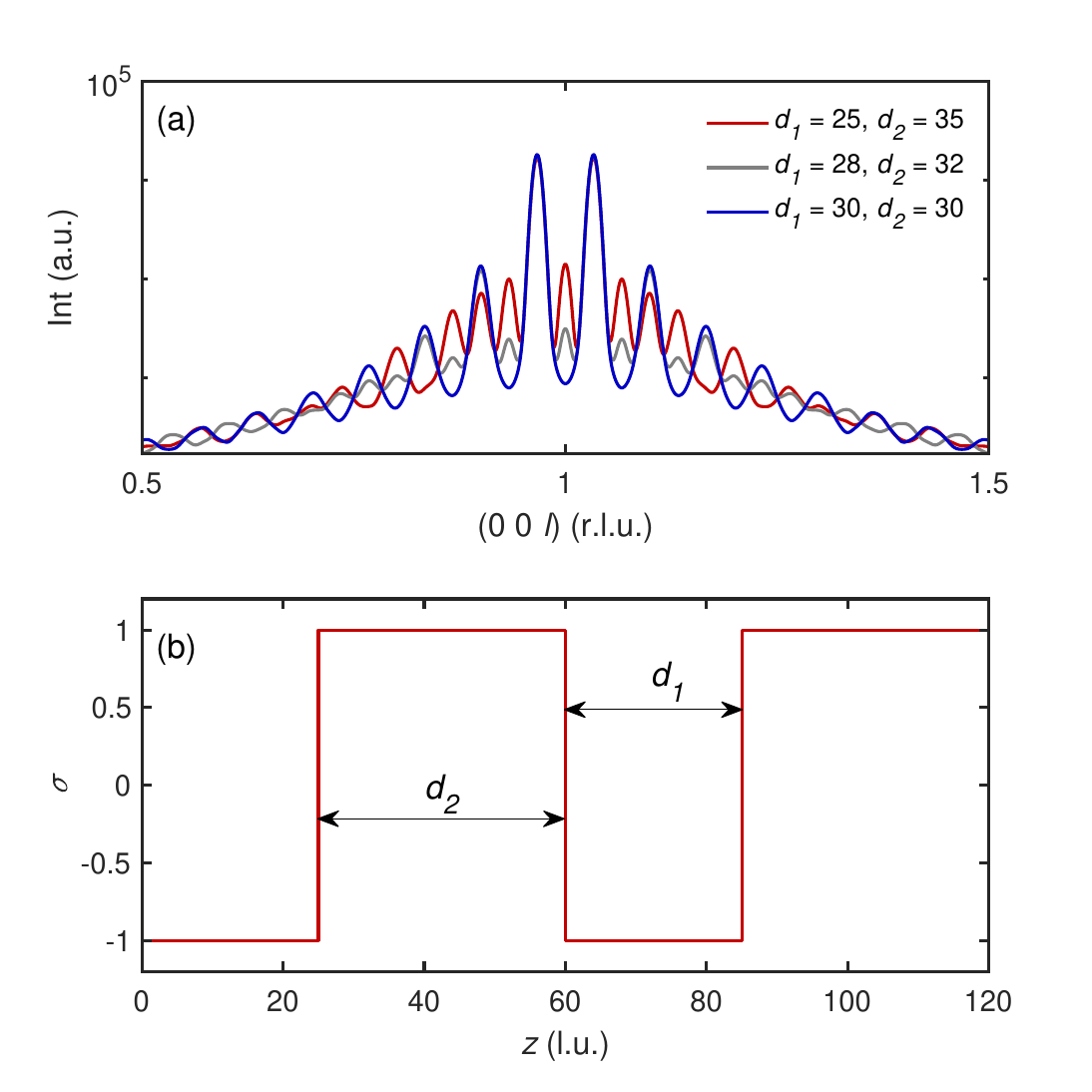}}
  \caption{~Temperature dependence of integrated magnetic intensity measured at different fields applied along $c$ axis.
  }
  \label{Fig_SM_different_domain_size}
  \vspace{-12pt}
\end{figure}

In the main article we demonstrate for $H_0 \neq 0, E_0 = 0$, as well as for $H_0 = 0, E_0 \neq 0$, the domain wall array remains equidistant and has a zero net polarization~(Fig.~4d). However, when both fields are non zero our theory predicts that the domain wall array become dimerized~(Fig.~4g). This behavior should strongly affect observed diffraction signal. To illustrate this we performed simple calculations assuming different degrees of dimerization as shown in Fig.~\ref{Fig_SM_different_domain_size}. 
For $d_1 = d_2$ the calculated curve consists of multiple peaks centered at odd harmonics of the primary propagation wavevector $\delta$. In contrast, when finite dimerization is present one can clearly see appearance of even-order harmonics along with the commensurate peak at (001). The intensity of the new peaks is controlled by the strength of dimerization.

\bibliography{bibliography}